\documentclass[aip,graphicx]{revtex4-1}

\usepackage{graphicx}
\usepackage{xcolor}
\usepackage{physics}
\usepackage[spanish,english]{babel}

\begin{document}


\title{Selective coupling and spectral evolution of topological domain-wall states under local perturbations}




\author{Ricardo Y. Díaz-Bonifaz}
\email[]{rdiazbnfz2015@ciencias.unam.mx}
\affiliation{Departamento de Física, Facultad de Ciencias, Universidad Nacional Autónoma de México}

\author{Edna M. Hernández}
\email[]{ednah@ciencias.unam.mx}
\thanks{Corresponding author}
\affiliation{Departamento de Física, Facultad de Ciencias, Universidad Nacional Autónoma de México}

\author{Carlos Ramírez}
\email[]{carlos@ciencias.unam.mx}
\thanks{Corresponding author}
\affiliation{Departamento de Física, Facultad de Ciencias, Universidad Nacional Autónoma de México}
\date{\today}

\begin{abstract}
In this work we investigate the spectral evolution of topological domain-wall states in a one-dimensional structure described by the Su-Schrieffer-Heeger (SSH) model. We show that a local potential applied to the domain wall induces a selective coupling between a topological state and specific bulk states. Moreover, by increasing the gate potential, we observe that the domain wall state energies asymptotically reach a bulk band, whereas new domain wall states appear on the opposite side of the band. As a consequence, the domain-wall-state spectrum evolves through a sequence of localized states that remain separated from the bulk continuum. These effects are also reflected in transport calculations, where the conductance peaks follow the evolution of the energy spectrum. To experimentally verify these phenomena, we realize an RLC topo-electrical circuit and perform impedance spectroscopy measurements. We find that the impedance matches the local density of states, exhibiting resonances at frequencies associated with the bulk energies that selectively couple to the domain wall topological state. 

\end{abstract}


\maketitle


\section{Introduction}
\label{sec:Introduction}

Topological invariants have been extensively studied in the last decades as a new paradigm to identify non-equivalent phases in condensed matter systems\cite{giustino_2021_2021}. A key feature of these non-conventional phases is the bulk-boundary correspondence. This principle establishes that a system with a non-trivial topological invariant hosts edge states that depend solely on the value of the invariant and are robust against adiabatic perturbations \cite{el-batanouny_advanced_2020}. Topological edge states (TES) are localized near the physical borders of the system as well as near boundaries between two non-equivalent topological phases, which we refer to as domain walls (DWs). These robust states have been proposed for promising applications such as dissipationless current pathways in low-dimensional circuits \cite{yasuda_quantized_2017,yin_valley_2024,hou_metallic_2020}, qubit formation \cite{kitaev_unpaired_2001,bomantara_quantum_2018,petropoulos_topological_2022}, quantum information transfer \cite{zurita_fast_2023,lang_topological_2017, kim_quantum_2021} and adiabatic charge pumping \cite{tapia-de-la-rosa_nanostructured_2024,longhi_topological_2019}. 

Topological edge states in 1D materials are particularly relevant for quantum computing applications. For instance, Zurita et al. proposed a quantum information transfer protocol based on a time-dependent Hamiltonian that varies adiabatically \cite{zurita_fast_2023}. In their scheme, an electron is transferred between the edges of a 1D structure exclusively through topological states localized in DWs. Since only few states participate in quantum transport, the protocol is expected to reduce decoherence and enhance fidelity. This idea builds on previous efforts to re-scale systems by forming DW lattices which, within a tight-binding (TB) framework, can be described  through re-scaled hopping parameters and on-site energies relative to the underlying lattice \cite{munoz_topological_2018}.

A key aspect of electron localization in topological DWs is that TES localized in neighboring domains are close to each other and might couple. This phenomenon has been studied using atomic-like effective models, which show that as TES approach to each other they form bonding-antibonding pairs with energies that differ from the original uncoupled TES \cite{han_topological_2023}. This effective model successfully explains why the number of edge states in a DW between Chern insulators with Chern numbers $C_{1}$ and $C_{2}$ is exactly  $|C_{2}-C_{1}|$. More recently, it has been shown that coupling between TES in infinite DWs in graphene nanoribbons within the quantum Hall regime produces a deformation of the Landau level flat bands, enabling phenomena such as anti-chirality, anomalous bulk-like dispersion and valley-specific current pathways \cite{diaz-bonifaz_band_2025}. These results demonstrate that TES couple when they are close to each other both spatially and energetically, which naturally raises the question if TES can couple with bulk states when brought into energetic proximity by a local potential. More generally, it is unclear how the energy spectrum evolves when a localized topological state approaches a bulk band, and whether this process can induce measurable rearrangements of the bulk spectrum itself.

Theoretical models of topological insulators (TIs) include the Su-Schrieffer-Heeger (SSH) \cite{su_solitons_1979}, Haldane \cite{haldane_model_1988} or Kane-Mele models \cite{kane_quantum_2005}. However, these models are often difficult to apply in real materials or require complex and expensive synthesis processes and experimental setups \cite{kong_opportunities_2011, ginley_topological_2016,he_review_2013}. Meta-material realizations such as cold atoms in optical lattices \cite{aidelsburger_artificial_2018,leder_real-space_2016}, phononic \cite{yin_valley_2024} and photonic \cite{cerjan_observation_2020} crystals or topo-electrical circuits \cite{lee_topolectrical_2018} are therefore powerful alternatives to observe or even utilize TES. Among these options, topo-electrical circuits provide a versatile and usually low-cost option for modelling topological systems. A simple way to emulate a 1D topological structure such as an SSH chain is by forming a lattice of RLC resonators with capacitive or inductive coupling \cite{liu_fully_2022, english_topologically_2026}. In this case, the circuit Laplacian is analogous to a TB Hamiltonian with a one-on-one relationship between frequency and energy, allowing to identify eigen-energies of the Hamiltonian as the frequencies for which there is a local maximum of impedance. These ideas were recently harnessed to develop a low-cost topoelectrical circuit that allows to simulate 1D topological insulators with a DW, gathering information of the energy spectrum through voltage measurements \cite{english_topologically_2026}.

In this work we investigate the coupling between TES in DWs and bulk states by modelling an SSH chain with a local potential applied to a DW. First, in section \ref{sec:Theory} we describe in detail the TB Hamiltonian of the SSH chain with DW and calculate the energy spectrum and wave functions, exposing phenomena related to the bulk-topological coupling. Section \ref{sec:Transport} explores the relation between these effects and transport properties. Finally, in Section \ref{sec:Circuit} we experimentally realize a topo-electrical analog of the SSH chain and compare impedance spectroscopy measurements with the theoretical local density of states (LDOS), obtaining direct experimental evidence of the selective bulk-topological coupling. For completeness, Appendix \ref{sec:LegoCircuit} summarizes a graph-based procedure that allows the construction of equivalent topo-electrical circuits from arbitrary tight-binding models.


\section{Selective coupling and spectral redistribution in a multi-domain SSH chain}
\label{sec:Theory}

\begin{figure}[h!]
\includegraphics[width=0.9\linewidth]{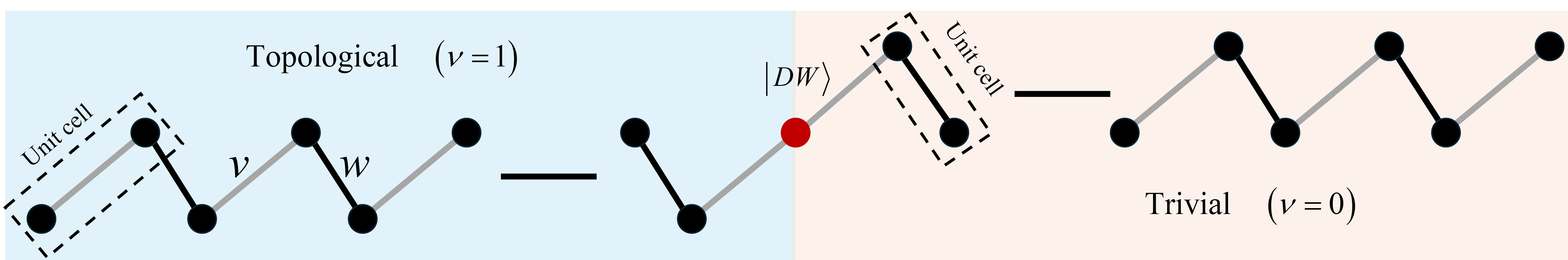}
\caption{\label{fig:SSH_Chain} Schematic representation of an SSH chain in which the interruption of the alternating pattern of the hopping parameters creates two non-equivalent domains. The hopping parameters $v$ and $w$ are represented by gray and black lines respectively. The red site, where the pattern is interrupted, is common between both domains and we call a domain wall (DW) site. The case shown corresponds to $|v|<|w|$, where a non-trivial domain appears in the left side and a trivial domain appears in the right side.}
\end{figure}

In the following we consider a dimerized chain described by a TB Hamiltonian with intra-cell hopping parameter $v$ and inter-cell hopping parameter $w$, which corresponds to the SSH model \cite{su_solitons_1979}. The topology of the chain is described by a winding number $\nu$ whose value depend on the relation between $|v|$ and $|w|$ \cite{asboth_short_2016}. When $|v|>|w|$ we have a trivial winding number $\nu=0$, while $|v|<|w|$ leads to a non-trivial topological phase as $\nu=1$. As proposed in Refs. \cite{munoz_topological_2018,zurita_fast_2023}, a multi-domain SSH chain can be built by interrupting the alternating pattern from the hoppings as shown in Fig. \ref{fig:SSH_Chain}, where we have repeated the inter-cell hopping $v$. For $|v|<|w|$ this creates a non-trivial domain in the left and a trivial domain in the right, where the last site in the left domain is the first site in the right domain. 

The idea represented in Fig. \ref{fig:SSH_Chain} can be extended to create a chain with $N_{DW}$ DWs created by pattern interruptions, where $N_{DW}+1$ domains alternate between non-trivial and trivial winding number.  The last site of every domain is the first site of the following domain. These common sites are the ones in which the pattern is interrupted and we refer to them as the DWs. Let us consider $N_{DW}+1$ domains, in each domain there are exactly $N$ hopping parameters with value $v$, but due to the alternation, this means that the domains alternate between $N$ and $N-1$ unit cells, where each cell has 2 sites (see Fig. \ref{fig:SSH_Chain}). Due to DWs being shared between domains, the total number of sites in the chain is $N_{sites} = (N_{DW}+1)(2N-1)+1$. If the sites in the chain are numbered from $0$ to $N_{sites}-1$,
the TB Hamiltonian can be written as

\begin{multline}
\label{eq:Hamiltonian}
    \hat{H}=\sum_{i=0}^{N_{sites}-2} \left(t_{i,i+1}\ket{i+1}\bra{i}+h.c.\right)+\sum_{i=1}^{N_{sites}-2}\left(\epsilon_{0}+\delta_{0,i(mod(2N-1))}\epsilon_{G}\right)\ket{i}\bra{i} \\ + (\epsilon_{0}+\delta^{GB}\epsilon_{G})\left(\ket{0}\bra{0} + \ket{N_{sites}-1}\bra{N_{sites}-1}\right).
\end{multline} 

The first sum in Eq. \ref{eq:Hamiltonian} term introduces the hopping parameters

\begin{equation} \label{eq:HoppingDef}
\begin{aligned}
    t_{i,i+1}=\frac{v+w}{2}+\frac{v-w}{2}(-1)^{i(mod(2N-1))},
\end{aligned}
\end{equation}
where $i(mod(2N-1))$ denotes the remainder of the integer division of $i$ over $2N-1$, resulting in the desired pattern for the chain with $N_{DW}$ DWs. A constant on-site energy $\epsilon_{0}$ is applied to all sites to introduce a rigid displacement of the chain energy spectrum \cite{pn-junctions}. The second term in Eq. \ref{eq:Hamiltonian} contains a Kronecker delta $\delta_{n,m}$ so that the additional on-site energy $\epsilon_{G}$ accounts for a gate potential applied to the DWs. This gate potential is  applied to the chain borders (sites $0$ and $N_{sites}-1$) depending on the value of $\delta^{GB}$, being  $\delta^{GB}=1$ for gated borders and $\delta^{GB}=0$ for ungated borders.

\begin{figure}
\includegraphics[width=1.\linewidth]{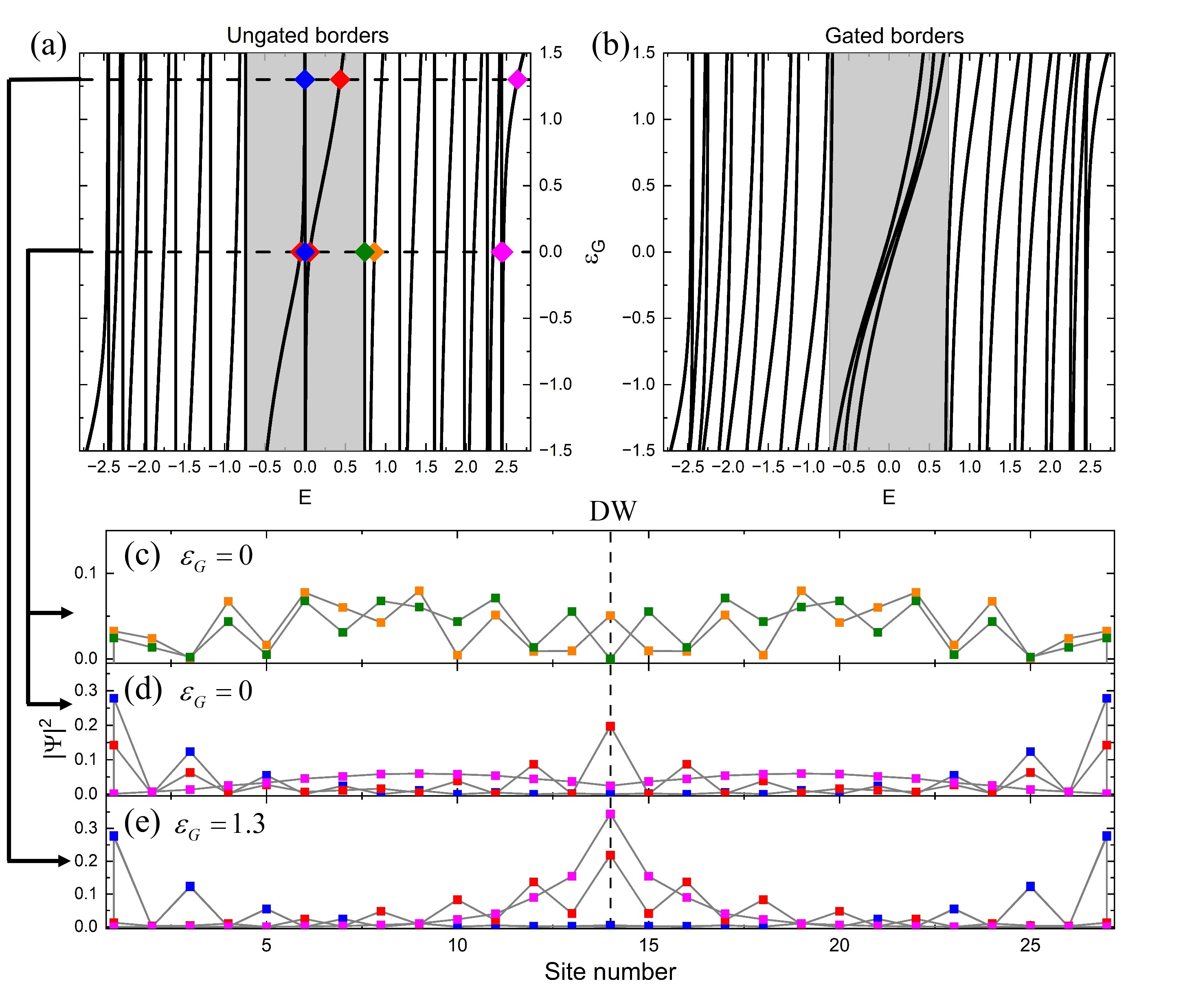}
\caption{\label{fig:Spectrum_WaveFunctions} (a) and (b) show the energy spectra for a multi-domain SSH chain with $N_{DW}=1$ DWs and $N=7$ unit cells per domain. The Hamiltonian parameters are $v=1$, $w=1.5$ and $\epsilon_{0}=0$. The spectra are calculated as $\epsilon_{G}$ is varied. Fig. (a) corresponds to a chain with ungated borders $(\delta^{GB}=0)$, while (b) corresponds to gated borders $(\delta^{GB}=1)$. The zones shaded in grey denote to the bulk band gap. We calculate the wave function for a selection of states from the chain with ungated borders and show their squared amplitudes $|\Psi|^2$ in (c-e). These selected states can be identified in the spectrum as colored dots that match the color of the wave function. The wave functions shown in (c) and (d) correspond to $\epsilon_{G}=0$ while the ones shown in (e) are calculated for $\epsilon_{G}=1.3$.}
\end{figure}

Let us consider a multi-domain SSH chain with $N_{DW}=1$ DWs and $N=7$, whose hopping parameters are $v=1$ and $w=1.5$. We consider a null constant on-site energy $\epsilon_{0}=0$. In Fig. \ref{fig:Spectrum_WaveFunctions} we show the energy spectrum of such chain as a function of the on-site energy $\epsilon_{G}$ that accounts for the gate potential. We consider the cases with ungated $(\delta^{GB}=0)$ borders, shown in Fig. \ref{fig:Spectrum_WaveFunctions}a, and gated borders $(\delta^{GB}=1)$, shown in Fig. \ref{fig:Spectrum_WaveFunctions}b. 

By construction, one of the two domains has non trivial topology, which leads to the appearance of topologically protected states in one end of the chain as well as in the DW. Moreover, as the structure is symmetric with respect to the DW, an additional localized state must appear in the other border of the chain regardless of the trivial topology of the domain \cite{zurita_fast_2023}. Thus, we expect 3 states with energies within the band gap, as can be seen in Figs. \ref{fig:Spectrum_WaveFunctions}a and \ref{fig:Spectrum_WaveFunctions}b, where the bulk band gap is shaded in gray. In the case with ungated borders, shown in Fig. \ref{fig:Spectrum_WaveFunctions}a, the gate potential is only applied to the DW and a single band gap state energy is modified as $\epsilon_{G}$ varies. This differs from the chain with gated borders as we can see that all three energies within the band gap change as $\epsilon_{G}$ varies. In both cases, these energies are always within the band-gap and never degenerate with bulk state energies but rather develop an asymptotic behavior near the bulk bands as $\epsilon_{G}$ gets further from $0$.

Let us now focus on the bulk states and define \textit{affected states} as those whose energies change as $\epsilon_{G}$ varies, while \textit{unaffected states} have energies that remain constant regardless of $\epsilon_{G}$. The energy spectrum of the chain with ungated borders, shown in Fig. \ref{fig:Spectrum_WaveFunctions}(a), is such that bulk states alternate between affected and unaffected states. When the band-gap energies approximate a bulk band, the affected bulk state energies increase in magnitude. Conversely, when the band-gap energies are further from a band, the affected state energies decrease in magnitude. This suggests a coupling between the band-gap states and the bulk states, forming bonding and anti-bonding pairs. Similar phenomena have already been observed for topological edge states (TES) from neighboring domains in multi-domain topological insulators, where the topological-topological coupling is promoted by reducing the spatial separation between TES \cite{han_topological_2023} or by adding intermediate domains \cite{diaz-bonifaz_band_2025}. The phenomena observed in Figs. \ref{fig:Spectrum_WaveFunctions}a and \ref{fig:Spectrum_WaveFunctions}b, however, differ from the topological-topological state coupling as it now involves topological and bulk states. Moreover, the edges and DWs are not spatially separated from the rest of sites from the chain, where the bulk-states have non-vanishing wave-functions. Thus, the bulk-topological coupling hereby presented is promoted by spectral proximity rather than spatial proximity. 

Since energy levels cannot cross, the unaffected states act as spectral barriers that confine the motion of the affected states. Consequently, the energy shifts induced by the DW potential are redistributed through a sequence of avoided crossings involving only the affected states. At the same time, the trace of the Hamiltonian increases linearly with the local potential, implying that the sum of all eigenvalues must increase as well. Since the unaffected states remain fixed and the remaining affected states are constrained by the spectral barriers, the corresponding spectral displacement must ultimately accumulate in the highest-energy bulk state for $\epsilon_G>0$ (or the lowest-energy bulk state for $\epsilon_G<0$), which is the only state not bounded by a spectral barrier on one side of the spectrum. As discussed below, this state progressively separates from the bulk band and acquires the character of a localized DW state.

The alternation between affected and unaffected states can be understood by analyzing their wave-functions, which are shown in Figs. \ref{fig:Spectrum_WaveFunctions}(c-e). For instance, Fig. \ref{fig:Spectrum_WaveFunctions}(c) presents the wave-functions of an unaffected and an affected bulk-state, shown in green and orange respectively. We can see that the unaffected wave function has a node in the DW site. This is not the case for the affected state, whose wave function does not vanish at the DW. Moreover, notice that both states have non-vanishing wave functions at the borders. This is consistent with the spectrum shown in Fig. \ref{fig:Spectrum_WaveFunctions}(b), where there are no unaffected bulk states, as their wave functions are non-zero in the borders, where the gate potential is also applied.

A similar analysis can be performed with the localized states within the band gap. Observe Fig. \ref{fig:Spectrum_WaveFunctions}(d), where we present in blue the wave function of the zero energy state and in red the rest of the band gap states for a null gate potential on-site energy $\epsilon_{G}=0$. Notice that the wave function of the zero energy state has maxima in the borders, while the remaining band gap states have maxima in both the borders and the DW. If we now consider $\epsilon_{G}=1.3$, as shown in Fig. \ref{fig:Spectrum_WaveFunctions}(e), we can see that the wave functions of the zero energy states remain unchanged but the wave function of the band gap state shown in red has now a single maximum in the DW. Thus, we see that the wave functions have maxima in both the DW and the borders when the band gap states have similar energies so they can couple. 

Another interesting state to compare between Figs. \ref{fig:Spectrum_WaveFunctions}(d) and \ref{fig:Spectrum_WaveFunctions}(e) is the highest energy state, shown in magenta. According to the spectral constraints discussed above, this state is the only affected bulk state that can accommodate the accumulated spectral displacement induced by the gate potential. Consequently, it plays a special role in the spectral evolution. Notice that for $\epsilon_{G}=0$ this state has a typical bulk wave function that extends over the entire chain. However, for $\epsilon_{G}=1.3$, this state no longer behaves as a typical bulk state and instead behaves like a DW localized state that decays into the bulk. An interesting feature of this newly created DW state is that it has an energy $E \approx 2.64$ that separates itself from the rest of the bulk state energies but it does not match the on-site energy $\epsilon_{G}=1.3$. The positive on-site energy can be understood as a repulsive potential in the DW that demands a greater energy for a particle to be localized in the DW, which is achieved by the coupling between bulk states and the band gap state with modified energy due to the $\epsilon_{G}$. Conversely, the lowest bulk state energy decreases for $\epsilon_{G}<0$. In this case, the gate potential is analogous to a potential well in the DW that allows a particle to be localized in the DW with a lower energy. Notice that in both the $\epsilon_{G}>0$ and $\epsilon_{G}<0$ scenarios the DW state energy does not match the local potential.

The resulting spectral evolution resembles a Newton's cradle, where the effect of the local perturbation is transferred through a sequence of spectrally constrained states and ultimately emerges in the outermost bulk state. In this analogy, the unaffected states act as spectral barriers that restrict the motion of the affected states, while the outermost state accumulates the net spectral displacement and eventually detaches from the band to form a new localized DW state.

\begin{figure}
\includegraphics[width=1.\linewidth]{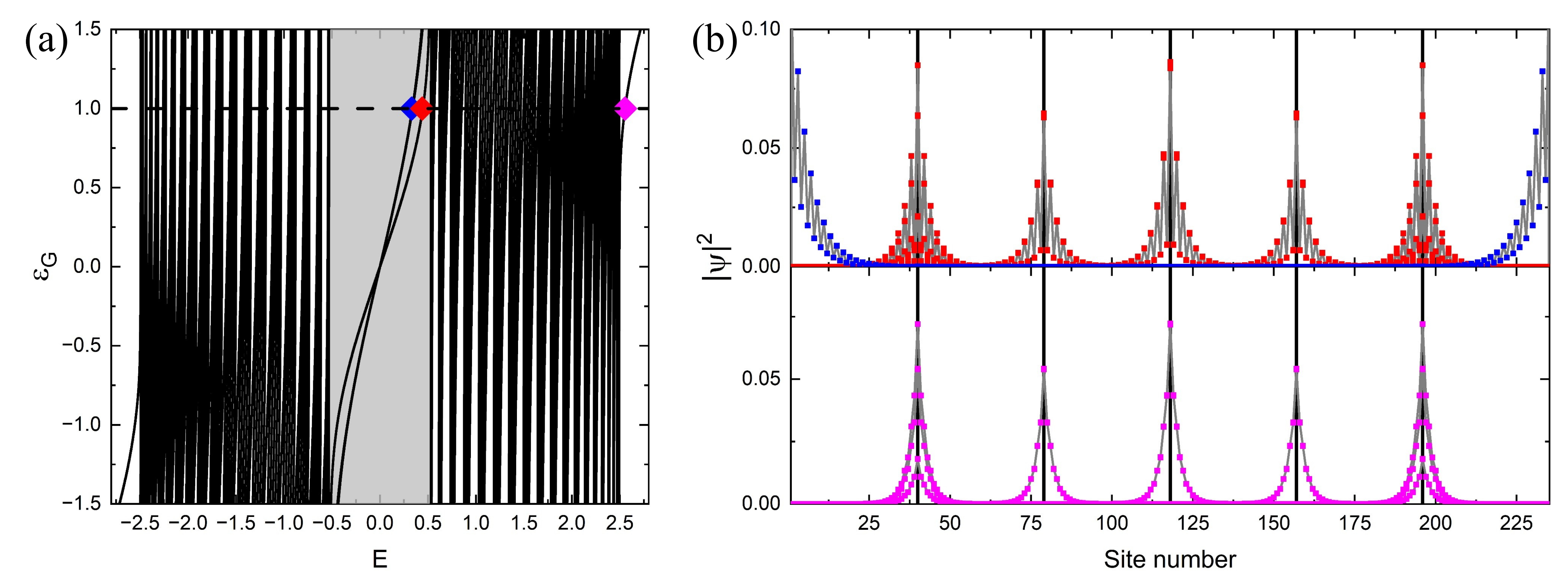}
\caption{\label{fig:BigChain_Spectrum} (a) shows the energy spectrum for a multi-domain SSH chain with $N_{DW}=5$ DWs and $N=20$ unit cells per domain as a function of $\epsilon_{G}$. The Hamiltonian parameters are $v=1$, $w=1.5$ and $\epsilon_{0}=0$. The zone shaded in grey is the bulk band gap. The colored dots indicate the states for which the wave functions shown in (b). The upper panel in (b) presents the squared amplitude $|\Psi|^2$ of band gap state wave functions, while the lower panel presents the wave function of the states with the highest energy.}
\end{figure}

A necessary question is whether the phenomena described from Fig. \ref{fig:Spectrum_WaveFunctions} are not due to the short length of the chain nor a particularity of a chain with a single DW. To discard these possibilities, we present in Fig. \ref{fig:BigChain_Spectrum}(a) the energy spectrum of a chain with $N=20$ and $N_{DW}=5$ as a function of $\epsilon_{G}$. As with the previous shorter chain, the Hamiltonian parameters are $v=1$, $w=1.5$ and $\epsilon_{0}=0$. This time, we consider the case with gated borders by making $\delta^{GB}=1$. Observe that once again, the band gap state energies vary as a function of $\epsilon_{G}$ and asymptotically approach the bulk bands. As the chain has 5 DWs and 2 edges, there are 7 states with energies within the band gap. As $\epsilon_{G}$ grows these 7 energies are divided in two groups, shown in red and blue in Fig. \ref{fig:BigChain_Spectrum}(b). This energy separation is intimately related with the wave-functions, as shown for $\epsilon_{G}=1$ in Fig. \ref{fig:BigChain_Spectrum}(b), where we can see that the lower energy states within the band gap are localized in the borders while the remaining states with greater energies have maxima in the DWs. Once again, there is a spectral separation between pure edge states and DW states despite the fact that the same gate potential is applied to both the edges and the DWs. Moreover, the highest energy states shown in magenta in Fig. \ref{fig:BigChain_Spectrum} have wave functions that are once again localized near the DWs excluding the borders despite the fact that the gate potential is applied to both the DWs and the borders. This fact further confirms that the appearance of these new DW states is not simply due to the repulsive potential in the DW but rather a consequence of the topological-bulk state coupling.

Moreover, as the number of sites between DWs increases, the bulk energy levels become more densely packed. Although the individual energy shifts of the affected states decrease, the same mechanism remains operative: the spectral displacement is redistributed throughout the band and ultimately accumulates at its edge. This suggests that the bulk density of states remains largely unchanged in the thermodynamic limit, while the emergence of the outermost localized DW state appears to be robust against increasing system size.

\section{Transport signatures of selective bulk-topological coupling} \label{sec:Transport}

The spectral evolution discussed above raises the question of how the selective bulk-topological coupling manifests in measurable electronic observables. One way of assessing this question is by analyzing the conductance $G$ of the multi-domain SSH chain. To calculate this, let us consider a finite multi-domain SSH chain described by the Hamiltonian shown in Eq. \ref{eq:Hamiltonian}. Let us now assume that each border is connected to a semi-infinite periodic one-dimensional chain with constant null on-site energy and constant hopping parameter $t_{C}$. We refer to these semi-infinite chains as \textit{auxiliary chains} and their main goal is to serve as waveguides that provide incoming and outgoing waves into the finite SSH chain. Under these conditions, the transmission function $T(E,\epsilon_{G})$ for an energy $E$ and a chain with gate on-site energy $\epsilon_{G}$ can be calculated using the recursive S-matrix method described in Ref.  \cite{RAMIREZ_SMat}. The transmission function $T(E,\epsilon_{G})$ is closely related to the electric conductance by the Landauer formula, that for zero temperature can be written as 

\begin{equation} \label{eq:conductance}
\begin{aligned}
    G(E,\epsilon_{G})=\frac{2e^2}{h}T(E,\epsilon_{G}),
\end{aligned}
\end{equation}
where $e$ is the electron charge and $h$ is Planck constant \cite{ryndyk_2016}.

\begin{figure*}
\includegraphics[width=1\linewidth]{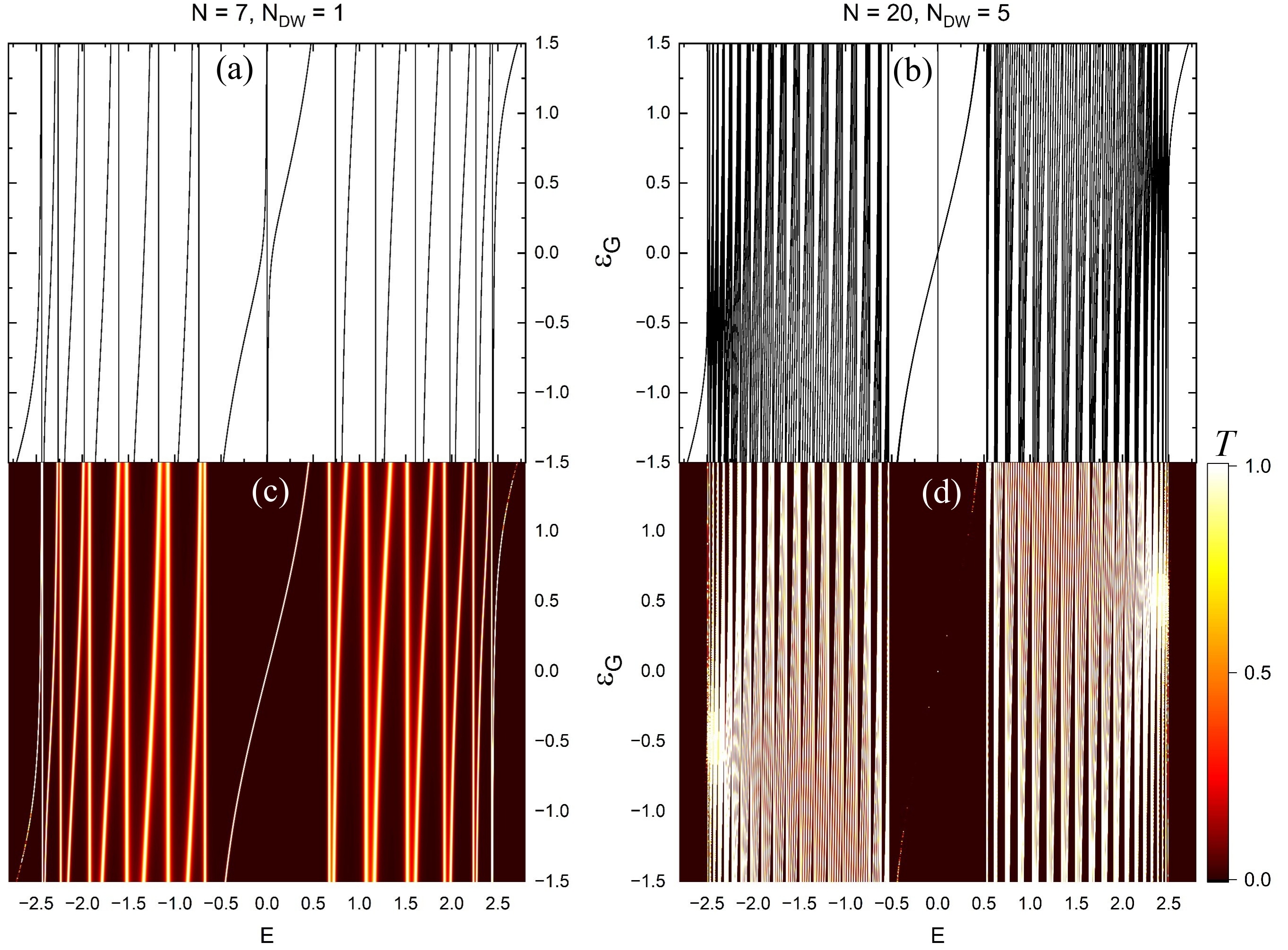}
\caption{\label{fig:Transport} Energy spectra of the multi-domain SSH chains for $(N=7, N_{DW}=1)$ in (a) and for $(N=20$, $N_{DW}=5$) in (b). Both cases correspond to ungated borders $\left(\delta^{GB} = 0 \right)$. In (c) and (d) we show the corresponding electric conductance $G$ as a function of both the energy $E$ and the gate potential on-site energy $\epsilon_{G}$. The conductance is shown in units of $G_{0}=\frac{2e^2}{h}$.}
\end{figure*}

Figure \ref{fig:Transport}(a-b) reproduces the energy spectra previously discussed for the short ($N=7$, $N_{DW}=1$) and long ($N=20$, $N_{DW}=5$) SSH chains with ungated borders. The corresponding conductance maps are shown in Figs. \ref{fig:Transport}(c) and \ref{fig:Transport}(d), allowing for a direct comparison between the spectral evolution and its transport signatures. By comparing the conductance maps with the corresponding energy spectra shown in Figs. \ref{fig:Transport}(a) and \ref{fig:Transport}(b), it can be seen that the conductance maxima closely follow the spectral branches discussed in the previous section. In particular, both the affected and unaffected bulk states can be identified in the conductance maps, demonstrating that the selective bulk-topological coupling is not restricted to the energy spectrum but also leaves observable signatures in transport quantities. However, it can also be seen in Fig. \ref{fig:Transport}c that the conductance in the short chain exhibits maxima for the band gap states as well, which does not occur for the long chain whose conductance is shown in Fig. \ref{fig:Transport}d. This is consistent with the fact that DW states decay exponentially into the bulk, which means that they are not able to contribute to the conductance if $N$ is big.

\section{Topo-electrical circuit meta-material realization}
\label{sec:Circuit}

The multi-domain SSH chains described in section \ref{sec:Theory} can be simulated by topo-electrical circuits following the steps described in Appendix \ref{sec:LegoCircuit}. Specifically, let us consider a chain with $N_{sites}=27$ sites in total. The goal of the circuit realization is not only to reproduce the SSH spectrum, but also to experimentally probe the signatures of the selective bulk-topological coupling identified in Section \ref{sec:Theory}. In particular, we seek to observe the selective response of the affected bulk states and the emergence of new domain-wall-localized states induced by the gate potential. 

As discussed in Appendix \ref{sec:LegoCircuit}, one possibility is to model each site as a node connected to a grounding inductor. We consider the same grounding inductor of $L=1\;\text{mH}$ for all sites. As shown in Fig. \ref{fig:laplacian_construction}a of Appendix A, we set a capacitive coupling between neighboring sites where the coupling capacitor alternates between $C_{1}=10\;\text{nF}$ and $C_{2}=47\;\text{nF}$. Currently, the proposed chain is analogous to the one shown in Fig. \ref{fig:laplacian_construction}b of Appendix A, whose Laplacian matrix has the form from Eqs. \ref{eq:Dim_Ham} with on-site energies and hopping parameters given by Eqs. \ref{eq:onsite_ens} and \ref{eq:hoppings} respectively. The on-site energy difference between the borders and the remaining sites of the chain can be removed by adding an additional capacitor $C_{2}$ that connects each border site to ground, resulting in a constant on-site energy $\epsilon_{0}=C_{1}+C_{2}$ over the entire chain. 

\begin{figure}
\centering
\includegraphics[width=1\linewidth]{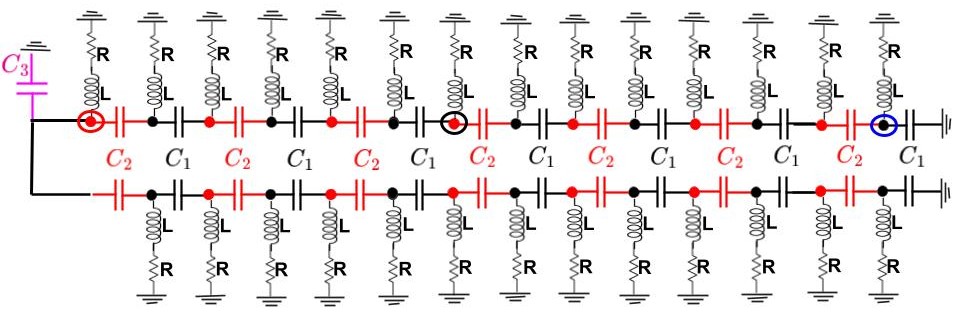}
\caption{
Schematic diagram of the experimental circuit with a domain wall, where $C_1=10 \text{ nF} \pm 10\%$, $C_2=47 \text{ nF}\pm 10\%$, $L=1
 \text{ mH}\pm 10\%$ and $R=1\; \Omega\pm 1\%$. The capacitor $C_{3}$ shown in magenta accounts for the gate potential on-site energy $\epsilon_{G}$ as described by Eq. \ref{eq:On-site_Gate}. The impedance spectroscopy measurements are done over 3 different sites which we refer to as the edge site ($n=0$, blue circle), bulk site ($n=7$, black circle) and domain wall ($n=13$, red circle). 
}\label{fig:Circuit_diagram}
\end{figure}

To create the DW, the alternating pattern in the hopping parameters is interrupted by coupling the $n=13$ site to both neighbors with a $C_{1}$ capacitor. This choice results in the desired hopping parameter pattern and we obtain a multi-domain SSH chain with $N_{DW}=1$ and $N=7$. It is worth noting that the pattern interruption in site $n=13$, which corresponds to the DW, results in an on-site energy of $2C_{1}$ that differs from the on-site energy $\epsilon_{0}$ of the remaining sites of the chain. We can interpret this as a gate potential induced on-site energy of $\epsilon_{G}=C_{1}-C_{2}$. We can further control the gate induced on-site energy by adding an additional grounding capacitor $C_{3}$ to the DW site. Thus, we are effectively modeling a chain with 

\begin{equation} \label{eq:On-site_Gate}
\begin{aligned}
    \epsilon_{G}=C_3+C_1-C_2,
\end{aligned}
\end{equation}
where the gate potential is only applied to the DW, that is, with $\delta^{GB}=0$. Finally, in order to successfully perform impedance spectroscopy measurements, a spectral broadening mechanism is introduced by adding $R=1\; \Omega \pm 1\%$ resistors in series with the grounding inductors. The resistors introduce a constant, pure imaginary on-site energy $\epsilon_{R}=\frac{i}{\omega R}$, making the Hamiltonian non-hermitian.
The complete diagram of the proposed circuit is shown in Fig. \ref{fig:Circuit_diagram}.

Let us provide a few remarks to further connect the circuit design with the experimental intent. First, the constant real term $C_{1}+C_{2}$ makes the entire energy spectrum positive. Second, notice that if $C_{3}=0$ we have $\epsilon_{G}<0$. It is shown in Section \ref{sec:Theory} that under these condition the lower energy bulk-band is where the affected and unaffected bulk states are spectrally separated. As we are working with a capacitive coupling rather than the inductive coupling proposed in Ref. \cite{english_topologically_2026}, the connection between frequency and energy is given as $E(\omega)=L^{-1}\omega^{-2}$, which means that in our equivalence between the circuit and the SSH chain the energy has units of capacitance. This choice yields three main advantages. First, the lower energy band is now associated with higher frequencies. Second, the $\omega^{-2}$ relation between frequency and energy widens the separation of the frequencies associated with low energy states. Third, the imaginary constant diagonal term $\epsilon_{R}$ is analogous to the artificial $i\eta$ term added to the energy denominator in the Green's function formalism that widens the LDOS \cite{datta1995electronic}. We exploit this fact by directly comparing the impedance spectroscopy with the theoretical LDOS of the multi-domain SSH chain.

The circuit is analyzed by measuring the node-to-ground impedance in the 3 nodes circled in Fig. \ref{fig:Circuit_diagram}. We respectively call these nodes the \textit{edge site} (site 0, blue line), the \textit{bulk site} (site 7, black line) and the \textit{domain wall} or \textit{DW site} (site 13, red line). The node-to-ground impedance was evaluated by placing a reference resistor $R_{ref}=220 \;\Omega \pm 10\%$ in series with the node of interest, effectively forming a simple voltage divider where the local impedance acts as the output impedance. By feeding monochromatic signals of amplitude $V_{in}$ and frequency $\omega$ with a function generator (Tektronix AFG320), we measure the node-to-ground voltage amplitude $V_{out}$ and phase angle $\phi$ with respect to the input using an oscilloscope (Rigol DH924). By defining $H_{V}=\frac{V_{out}}{V_{in}}$, the impedance amplitude $|Z_n(\omega)|$ of the $n-$th node can be calculated as

\begin{equation} \label{eq:imp_exp}
\begin{aligned}
    |Z_{n}(\omega)|=\frac{R_{ref}}{|H_{V}^{-1}e^{-i \phi}-1|}.
\end{aligned}
\end{equation}

\begin{figure*}
\includegraphics[width=1\linewidth]{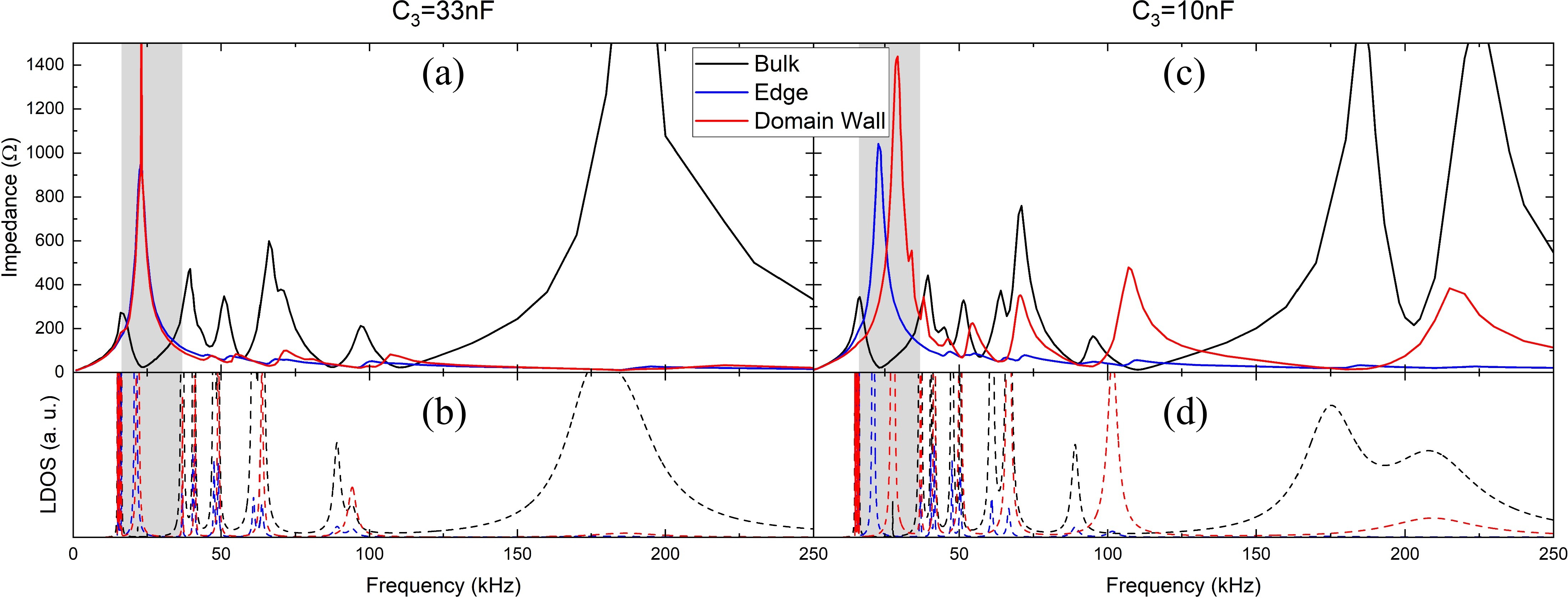}
\caption{\label{fig:Impedances} In (a) and (c) we show the impedance as a function of frequency measured in 3 different nodes of the circuit shown in Fig. \ref{fig:Circuit_diagram}. We refer to these sites as the \textit{edge site} (site 0, blue line), the \textit{bulk site} (site 7, black line) and the \textit{domain wall} or \textit{DW site} (site 13, red line). The color code of the impedance spectroscopy lines matches the color of the circled nodes of the circuit shown in Fig. \ref{fig:Circuit_diagram}. In (b) and (d) we show the theoretical LDOS of the multi-domain SSH chain calculated from Eq. \ref{eq:LDOS}. In (a) and (b) we consider a chain with $C_{3}=33\text{ nF}$ while in (c) and (d) we have $C_{3}=10\text{ nF}$. The LDOS is shown in arbitrary units. In all cases, the gray shaded area indicate the frequency range associated with the bulk band gap.}
\end{figure*}

The impedance spectroscopy measurements are shown in Fig. \ref{fig:Impedances} (a) and (c) for the circuit shown in Fig. \ref{fig:Circuit_diagram} with $C_{3}=33 \text{ nF} \pm 10\%$ and $C_{3}=10\text{ nF} \pm 10\%$, respectively. 

To address the relation between the impedance and energy spectrum of the SSH chain, we calculate the theoretical LDOS of the $n-$th node using the Green's function formalism as 

\begin{equation} \label{eq:LDOS}
\begin{aligned}
    \text{LDOS}_{n}(E)=-\frac{1}{\pi} \text{Im} (G_{n,n}^{R}(E)),
\end{aligned}
\end{equation}
where $G_{n,n}^{R}(E)$ is the $n-$th diagonal term of the retarded Green's function matrix $\mathbf{G}^{R}(E)=\left[(E+i \eta) \mathbf{I} - \mathbf{H} \right]^{-1}$ with $\mathbf{H}$ being the Hamiltonian matrix and $\mathbf{I}$ being the identity matrix \cite{ryndyk_2016}. The term $\eta=1e^{-8}$ is a small, positive number that introduces an artificial non-hermiticity that broadens the LDOS peaks. The resulting LDOS are shown in Figs. \ref{fig:Impedances} (b) and (d) when considering $C_{3}=33 \text{ nF}$ and $C_{3}=10 \text{ nF}$ respectively.

Consider the impedance curves from Fig. \ref{fig:Impedances}(a), which correspond to the circuit shown in Fig. \ref{fig:Circuit_diagram} with $C_{3}=33 \text{ nF}$. In the equivalent SSH chain, this leads to a gate potential on-site energy of $\epsilon_{G}=-4\text{ nF}$. Observe that the impedance from both the edge site (blue curve) and the DW site (red curve) exhibit a maximum within the band gap and decay in the bulk bands, as expected for the topological edge and DW states. This behaviour is different from the impedance measured in the bulk site (black line), which has no maxima within the band gap. Moreover, the maxima of the bulk site impedance are in agreement with the maxima of the LDOS shown in Fig. \ref{fig:Impedances}(b), which was calculated for the exact same chain. These results can be compared with the ones from Figs. \ref{fig:Impedances}(c) and (d), for which $C_{3}=10 \text{ nF}$. In this case we have $\epsilon_{G}=-27\text{ nF}$, which means that it has a larger gate potential applied over the DW compared to the chain with $C_{3}=33 \text{ nF}$. Once again, we observe that the edge and DW node impedances have maxima within the band gap and that the bulk site impedance maxima match the bulk LDOS maxima. However, notice that in this case we have that the band gap maxima of the edge (blue) and DW (red) site impedances are separated and match the theoretical LDOS. This is consistent with the fact that a greater gate potential is applied on the DW. Another interesting feature is that the impedance measured in the DW site exhibits greater maxima within the bands, matching the theoretical LDOS. Moreover, these maxima do not appear for every bulk energy and instead alternate, matching the selection rule identified in Section \ref{sec:Theory}. Only the bulk states that participate in the selective bulk-topological coupling produce a strong impedance response at the DW site. This constitutes direct experimental evidence of the affected and unaffected states predicted by the spectral analysis.

\begin{figure*}
\includegraphics[width=1\linewidth]{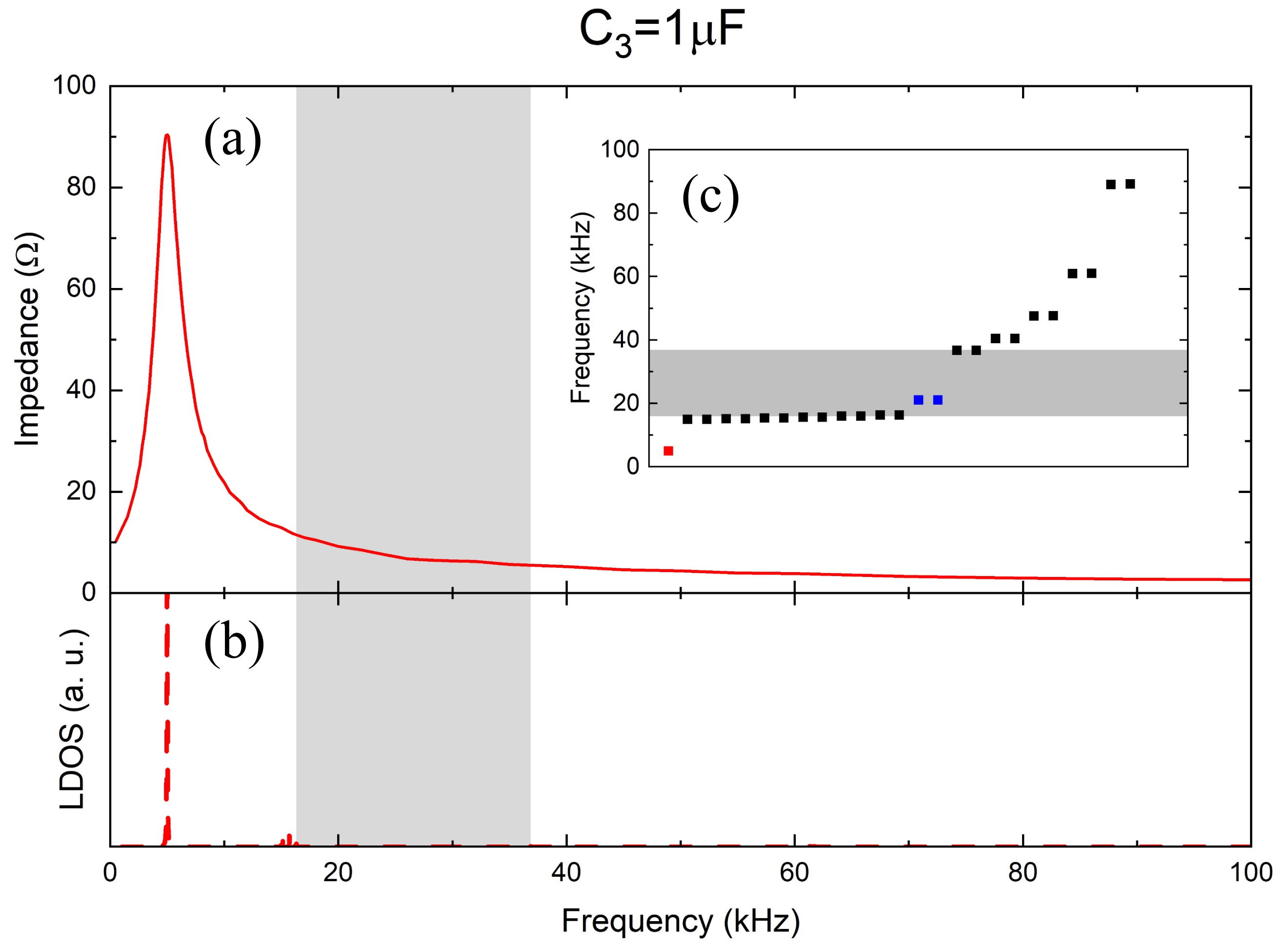}
\caption{\label{fig:1muF_Capacitor} (a) Impedance measured in the domain wall site (site 13) as a function of the frequency for the circuit shown in Fig. \ref{fig:Circuit_diagram} with $C_{3}=1 \;\mu F$. In (b) we show the theoretical LDOS calculated for the equivalent multi-domain SSH chain. The frequencies associated with all the Hamiltonian eigenvalues are shown in (c), with the edge states shown in blue and the DW state shown in red. The gray shaded zones in (b) and (c) denote the frequency range associated with the bulk band gap.}
\end{figure*}

Another phenomenon found in the theoretical analysis made in Section \ref{sec:Theory} is that, when a big gate potential is applied to the DW, the highest or lowest energy in the spectrum increases in magnitude, separating itself from the bulk band, and the corresponding state is localized near the DWs. This can be seen in the magenta wave functions shown in Figs. \ref{fig:Spectrum_WaveFunctions} and \ref{fig:BigChain_Spectrum}. This fact can also be observed experimentally using the circuit shown in Fig. \ref{fig:Circuit_diagram} by setting $C_{3}=1 \; \mu F$. Notice from Eq. \ref{eq:On-site_Gate}, that this induces a big on-site term $\epsilon_{G}=0.963 \;\mu F$. As the on-site energy is positive, we expect the highest energy state to behave as a DW state. As we have $E \propto \omega^{-2}$, this state will have the lowest frequency. A direct diagonalization of the Hamiltonian from Eq. \ref{eq:Hamiltonian} leads to the frequency spectrum shown in Fig. \ref{fig:1muF_Capacitor}(c), where we can see that there are only two band gap states (blue), which are topological edge states. Additionally, we can see that the lowest frequency (red) is far from the rest of bulk states. We can confirm that this state is localized in the DW by performing the impedance spectroscopy in the DW site, which leads to the impedance presented in Fig. \ref{fig:1muF_Capacitor}(a). Notice that there is a single maximum that lies well beyond the band gap and matches the LDOS calculated for the exact same site. This result is consistent with the spectral evolution discussed in Section \ref{sec:Theory}, where the accumulated spectral displacement is transferred to the outermost bulk state, causing it to detach from the band and acquire the character of a localized DW state.

\section{Conclusions}

In this work we investigated the spectral evolution of topological domain-wall (DW) states in a multi-domain one-dimensional topological insulator subjected to a local gate potential. We found that the local potential induces a selective coupling between DW and bulk states, giving rise to two distinct classes of bulk states: affected states, whose energies evolve with the gate potential, and unaffected states, whose wave functions possess nodes at the DW sites. The unaffected states act as spectral barriers that constrain the motion of the affected states through a sequence of avoided crossings. As a consequence, the spectral displacement induced by the local potential accumulates at the edge of the bulk band, where the highest-energy state (for $\epsilon_G>0$) or lowest-energy state (for $\epsilon_G<0$) separates from the bulk continuum and progressively acquires the character of a localized DW state. Thus, the spectral evolution is analogous to a transfer of the DW-state localization from the band gap to the opposite side of the bulk band.

We further showed that these phenomena leave observable signatures in quantum transport. In particular, the conductance spectrum reproduces the spectral branches identified in the energy analysis, allowing the affected and unaffected states to be distinguished directly from transport measurements. The conductance maxima follow the same spectral evolution observed in the energy spectrum, demonstrating that the selective bulk-topological coupling is not merely a spectral effect but also manifests in measurable transport observables.

The theoretical predictions were experimentally verified through a topo-electrical RLC circuit realization of the multi-domain SSH chain. Impedance spectroscopy measurements revealed resonances that accurately reproduce the theoretical local density of states and, more importantly, distinguish between affected and unaffected bulk states. In this way, the topo-electrical platform provides direct experimental evidence of the selective bulk-topological coupling. Furthermore, the emergence of a localized DW state outside the original band gap was experimentally observed, confirming the spectral evolution predicted by the theoretical analysis.

The analysis presented here suggests that selective bulk-topological coupling and the resulting spectral evolution may constitute a general mechanism in systems where localized topological states coexist with extended bulk states and local perturbations. Although demonstrated here for the SSH model, similar phenomena may arise in alternative one-dimensional topological systems such as Kitaev chains \cite{kitaev_unpaired_2001}, as well as in higher-order topological insulators hosting zero-dimensional topological states \cite{Benalcazar}. More generally, the notion of spectral barriers introduced in this work provides a useful framework to understand how local perturbations are redistributed through the spectrum of topological systems. Finally, the graph-based mapping summarized in Appendix \ref{sec:LegoCircuit} offers a practical route for implementing and experimentally exploring these phenomena in topo-electrical platforms.

\begin{acknowledgments}
This work was supported by UNAM-PAPIIT IN116025 and Secihti-Ciencia de Frontera CF-2023-1-136. Computations were performed at Miztli under project LANCAD-UNAM-DGTIC-329. Ricardo Y. Diaz-Bonifaz thanks SECIHTI for the graduate scholarship granted.
\end{acknowledgments}

\appendix
\section{Graph representation of topoelectrical circuits}
\label{sec:LegoCircuit}

In this appendix we introduce a graphical representation that allows the circuit Laplacian to be constructed directly from local circuit elements. Instead of writing Kirchhoff equations node by node, each element is associated with a local graph contribution. The complete circuit Laplacian is then obtained through the superposition of these elementary building blocks.

\subsection{Elementary graph rules}

\begin{figure}[t]
\centering
\includegraphics[width=0.5\columnwidth]{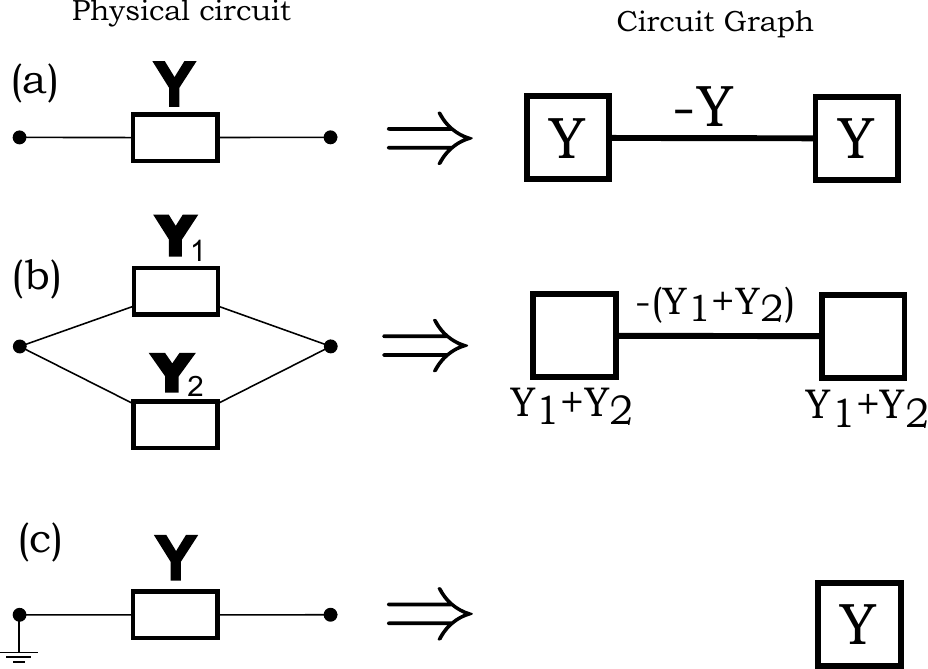}
\caption{
Elementary rules of the circuit graph representation.
(a) Generic two-terminal element of admittance $Y$.
(b) Parallel elements combine through the sum of their admittances.
(c) Grounded element.
}
\label{fig:graph_rules}
\end{figure}

The basic construction rules are summarized in Fig.~\ref{fig:graph_rules}. Consider first an arbitrary two-terminal circuit element characterized by an admittance $Y(\omega)$, as shown in Fig.~\ref{fig:graph_rules}(a). In the circuit graph representation, each node receives a local contribution equal to $Y$, while the connection between the two nodes contributes $-Y$. The corresponding matrix contribution is therefore

\begin{equation}
\Delta J=
\begin{pmatrix}
Y & -Y\\
-Y & Y
\end{pmatrix}.
\label{eq:elementary_block}
\end{equation}
This elementary block constitutes the fundamental building unit of the graphical construction.

Because the circuit Laplacian is a linear operator, the contributions of different elements add directly. As an immediate consequence, parallel elements combine through the sum of their admittances. Figure~\ref{fig:graph_rules}(b) illustrates the case of two elements connected in parallel, leading to an effective admittance

\begin{equation}
Y_{\rm eff}=Y_1+Y_2.
\end{equation}
The corresponding graph is therefore identical to that of a single element with admittance $Y_{\rm eff}$.

Grounded elements are treated similarly. Since the ground node does not constitute a dynamical degree of freedom, it is removed from the graph while retaining its local contribution. Consequently, the grounded element shown in Fig.~\ref{fig:graph_rules}(c) produces only a diagonal contribution,

\begin{equation}
\Delta J=(Y).
\label{eq:ground_contribution}
\end{equation}

Finally, larger networks can be assembled by combining the elementary operations described above. Figure~\ref{fig:graph_rules}(d) illustrates a series connection of two admittances $Y_1$ and $Y_2$. Observe that the diagonal element associated with a given node is obtained from the sum of all admittances incident on that node, whereas the off-diagonal element connecting two nodes equals minus the admittance directly linking them.

The graphical rules introduced above are independent of the physical nature of the circuit element. Different implementations simply correspond to different admittances and therefore generate circuit graphs with the same structure but different edge weights.

For a capacitor, an inductor, and a resistor, the corresponding
admittances are

\begin{equation}
Y_C=i\omega C,
\qquad
Y_L=\frac{1}{i\omega L},
\qquad
Y_R=\frac{1}{R}.
\end{equation}
Consequently, capacitive, inductive, and resistive couplings are represented by exactly the same graph topology, differing only in the admittance assigned to each graph element. More complex circuit components or combinations of elements can be incorporated straightforwardly through their equivalent admittance and the superposition rules described above.

\subsection{Construction of the circuit Laplacian}

Once the elementary graph rules have been established, arbitrary circuits can be translated directly into graph form. The complete circuit Laplacian is obtained by summing the local contributions associated with all circuit elements,

\begin{equation}
J(\omega)
=
\sum_{\alpha}
\Delta J_{\alpha}.
\label{eq:laplacian_sum}
\end{equation}

An additional feature of the circuit-graph representation is that it is formulated directly in terms of the dynamical degrees of freedom of the system. Ground nodes are eliminated through the rule of Fig.~\ref{fig:graph_rules}(c), while their contributions are retained as local graph terms. Furthermore, the graph is organized according to the structure of the circuit Laplacian itself: graph vertices contribute to diagonal matrix elements, whereas graph edges contribute to off-diagonal ones. The circuit graph can therefore be viewed as an intermediate representation connecting the physical circuit to the corresponding Laplacian matrix.

\begin{figure}[t]
\centering
\includegraphics[width=0.5\columnwidth]{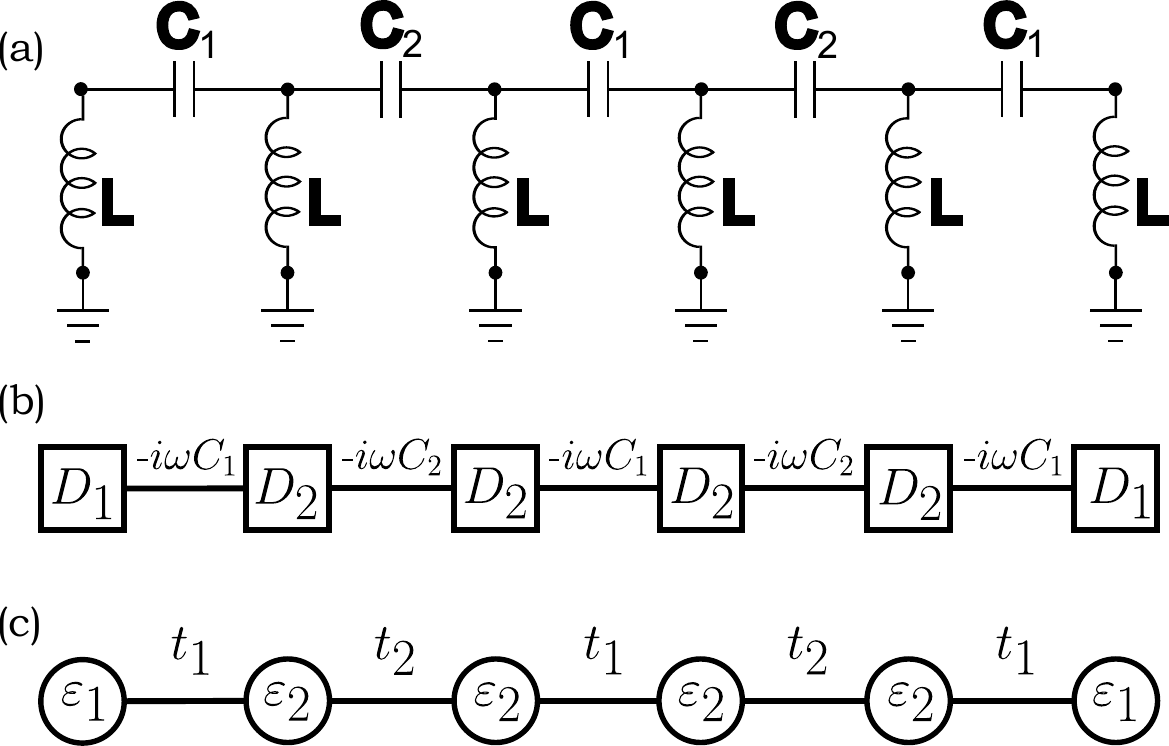}
\caption{
(a) Dimerized capacitive circuit.
(b) Corresponding circuit graph.
(c) Effective tight-binding graph.
}
\label{fig:laplacian_construction}
\end{figure}

As an illustrative example, Fig.~\ref{fig:laplacian_construction}(a) shows a dimerized capacitive circuit. Applying the rules introduced in Fig.~\ref{fig:graph_rules}, the circuit can be translated directly into the graph shown in Fig.~\ref{fig:laplacian_construction}(b). The diagonal contributions $D_1$ and $D_2$ are obtained from the sum of all admittances incident on the corresponding graph vertices,

\begin{equation}
\begin{aligned}
D_1 &= i\omega C_1+\frac{1}{i\omega L},\\
D_2 &= i\omega C_1+i\omega C_2+\frac{1}{i\omega L},
\end{aligned}
\label{eq:D12}
\end{equation}
while each graph edge contributes the negative of the admittance associated with the corresponding capacitor. The resulting graph therefore contains all the information required to construct the circuit Laplacian, which takes the form

\begin{equation}
J(\omega)=
\begin{pmatrix}
D_1 & -i\omega C_1 & 0 & 0 & \cdots\\
-i\omega C_1 & D_2 & -i\omega C_2 & 0 & \cdots\\
0 & -i\omega C_2 & D_2 & -i\omega C_1 & \cdots\\
0 & 0 & -i\omega C_1 & D_2 & \cdots\\
\vdots & \vdots & \vdots & \vdots & \ddots
\end{pmatrix}.
\label{eq:dimerized_laplacian}
\end{equation}

The alternating capacitive couplings are directly reflected in the off-diagonal structure of Eq.~(\ref{eq:dimerized_laplacian}), whereas the diagonal entries are determined entirely by the local graph contributions. The graph therefore provides an intermediate representation connecting the physical circuit topology to the corresponding Laplacian matrix. The same construction can now be used to establish the connection with an effective tight-binding description.

\subsection{Effective tight-binding representation}

The graph representation introduced above provides a direct construction of the circuit Laplacian $J(\omega)$. In order to establish a connection with tight-binding models, it is convenient to identify a global frequency-dependent factor $Y_0(\omega)$ and rewrite the Laplacian as

\begin{equation}
J(\omega)
=
Y_0(\omega)
\left[
H_{\rm eff}(\omega)-E(\omega)I
\right].
\label{eq:tb_mapping}
\end{equation}

Here, $H_{\rm eff}$ defines an effective tight-binding Hamiltonian, while $E(\omega)$ plays the role of the spectral parameter. The choice of $Y_0(\omega)$ and $E(\omega)$ is generally not unique. Different normalizations may be adopted depending on the particular features one wishes to emphasize in the effective description, leading to alternative effective representations of the same circuit Laplacian.

For the circuit graph shown in Fig.~\ref{fig:laplacian_construction}(b), the off-diagonal elements of Eq.~(\ref{eq:dimerized_laplacian}) are proportional to $i\omega$. It is therefore convenient to factor out this common term by choosing
\[
Y_0(\omega)=i\omega.
\]
The remaining diagonal contribution associated with the grounded inductors can then be absorbed into the spectral parameter by choosing
\[
E(\omega)=\frac{1}{\omega^2L},
\]
one obtains the effective tight-binding graph shown in Fig.~\ref{fig:laplacian_construction}(c), corresponding to
\begin{equation}
    H_{\rm eff}=
    \begin{pmatrix}
    \varepsilon_1 & t_1 & 0 & 0 & \cdots\\
    t_1 & \varepsilon_2 & t_2 & 0 & \cdots\\
    0 & t_2 & \varepsilon_2 & t_1 & \cdots\\
    0 & 0 & t_1 & \varepsilon_2 & \cdots\\
    \vdots & \vdots & \vdots & \vdots & \ddots
    \end{pmatrix}.
    \label{eq:Dim_Ham}
\end{equation}
where the onsite energies become
\begin{equation}
    \varepsilon_1=C_1
    \qquad \text{and} \qquad
    \varepsilon_2=C_1+C_2,
    \label{eq:onsite_ens}
\end{equation}
while the hopping amplitudes are
\begin{equation}
    t_1=-C_1
    \qquad \text{and} \qquad
    t_2=-C_2.
    \label{eq:hoppings}
\end{equation}
The particular choice $Y_0(\omega)=i\omega$ and $E(\omega)=1/(\omega^2L)$ is especially convenient because it produces a Hermitian tight-binding Hamiltonian with frequency-independent parameters. However, this is not a requirement of the formalism. Alternative choices may lead to effective descriptions with frequency-dependent on-site energies and hopping amplitudes or, more generally, to non-Hermitian effective Hamiltonians. In this sense, the effective description inherits some degree of flexibility from the choice of normalization used to separate the frequency dependence of the circuit Laplacian. Similar energy-dependent effective parameters are commonly encountered in effective tight-binding descriptions obtained through decimation or renormalization procedures.\cite{Farchioni1992}

Resistive elements naturally generate imaginary contributions to the effective parameters, which within the tight-binding language correspond to non-Hermitian corrections describing dissipation and spectral broadening.


%
%

%


\bibliography{RefsTopo2}

@article{yasuda_quantized_2017,
	title = {Quantized chiral edge conduction on domain walls of a magnetic topological insulator},
	volume = {358},
	url = {https://www.science.org/doi/10.1126/science.aan5991},
	doi = {10.1126/science.aan5991},
	number = {6368},
	urldate = {2025-02-21},
	journal = {Science},
	publisher = {American Association for the Advancement of Science},
	author = {Yasuda, K. and Mogi, M. and Yoshimi, R. and Tsukazaki, A. and Takahashi, K. S. and Kawasaki, M. and Kagawa, F. and Tokura, Y.},
	month = dec,
	year = {2017},
	pages = {1311--1314},
}

@article{yin_valley_2024,
	title = {Valley edge states as bound states in the continuum},
	volume = {69},
	issn = {20959273},
	url = {https://linkinghub.elsevier.com/retrieve/pii/S2095927324002251},
	doi = {10.1016/j.scib.2024.04.007},
	language = {english},
	number = {11},
	urldate = {2025-02-21},
	journal = {Science Bulletin},
	author = {Yin, Shunda and Ye, Liping and He, Hailong and Huang, Xueqin and Ke, Manzhu and Deng, Weiyin and Lu, Jiuyang and Liu, Zhengyou},
	month = jun,
	year = {2024},
	pages = {1660--1666},
}

@article{han_topological_2023,
	title = {Topological junctions in high-{Chern}-number quantum anomalous {Hall} systems},
	volume = {108},
	url = {https://link.aps.org/doi/10.1103/PhysRevB.108.115302},
	doi = {10.1103/PhysRevB.108.115302},
	number = {11},
	urldate = {2025-05-09},
	journal = {Physical Review B},
	publisher = {American Physical Society},
	author = {Han, Yulei and Pan, Shiyao and Qiao, Zhenhua},
	month = sep,
	year = {2023},
	pages = {115302},
}

@article{hou_metallic_2020,
	title = {Metallic network of topological domain walls},
	volume = {101},
	url = {https://link.aps.org/doi/10.1103/PhysRevB.101.201403},
	doi = {10.1103/PhysRevB.101.201403},
	number = {20},
	urldate = {2025-05-09},
	journal = {Physical Review B},
	publisher = {American Physical Society},
	author = {Hou, Tao and Ren, Yafei and Quan, Yujie and Jung, Jeil and Ren, Wei and Qiao, Zhenhua},
	month = may,
	year = {2020},
	pages = {201403},
}

@article{giustino_2021_2021,
	title = {The 2021 quantum materials roadmap},
	volume = {3},
	issn = {2515-7639},
	url = {https://dx.doi.org/10.1088/2515-7639/abb74e},
	doi = {10.1088/2515-7639/abb74e},
	language = {english},
	number = {4},
	urldate = {2025-05-09},
	journal = {Journal of Physics: Materials},
	publisher = {IOP Publishing},
	author = {Giustino, Feliciano and Lee, Jin Hong and Trier, Felix and Bibes, Manuel and Winter, Stephen M and Valentí, Roser and Son, Young-Woo and Taillefer, Louis and Heil, Christoph and Figueroa, Adriana I and Plaçais, Bernard and Wu, QuanSheng and Yazyev, Oleg V and Bakkers, Erik P A M and Nygård, Jesper and Forn-Díaz, Pol and De Franceschi, Silvano and McIver, J W and Torres, L E F Foa and Low, Tony and Kumar, Anshuman and Galceran, Regina and Valenzuela, Sergio O and Costache, Marius V and Manchon, Aurélien and Kim, Eun-Ah and Schleder, Gabriel R and Fazzio, Adalberto and Roche, Stephan},
	month = jan,
	year = {2021},
	pages = {042006},
}

@book{el-batanouny_advanced_2020,
	address = {Cambridge},
	title = {Advanced {Quantum} {Condensed} {Matter} {Physics}: {One}-{Body}, {Many}-{Body}, and {Topological} {Perspectives}},
	isbn = {978-1-108-48084-0},
	shorttitle = {Advanced {Quantum} {Condensed} {Matter} {Physics}},
	url = {https://www.cambridge.org/core/books/advanced-quantum-condensed-matter-physics/DED4815FE0BF969C02660325B8FAFF95},
	doi = {10.1017/9781108691291},
	urldate = {2025-05-30},
	publisher = {Cambridge University Press},
	author = {El-Batanouny, Michael},
	year = {2020},
}

@article{zurita_fast_2023,
	title = {Fast quantum transfer mediated by topological domain walls},
	volume = {7},
	url = {https://quantum-journal.org/papers/q-2023-06-22-1043/},
	doi = {10.22331/q-2023-06-22-1043},
	language = {english},
	urldate = {2025-05-30},
	journal = {Quantum},
	publisher = {Verein zur Förderung des Open Access Publizierens in den Quantenwissenschaften},
	author = {Zurita, Juan and Creffield, Charles E. and Platero, Gloria},
	month = jun,
	year = {2023},
	pages = {1043},
}

@article{haldane_model_1988,
	title = {Model for a {Quantum} {Hall} {Effect} without {Landau} {Levels}: {Condensed}-{Matter} {Realization} of the "{Parity} {Anomaly}"},
	volume = {61},
	copyright = {http://link.aps.org/licenses/aps-default-license},
	issn = {0031-9007},
	shorttitle = {Model for a {Quantum} {Hall} {Effect} without {Landau} {Levels}},
	url = {https://link.aps.org/doi/10.1103/PhysRevLett.61.2015},
	doi = {10.1103/PhysRevLett.61.2015},
	language = {english},
	number = {18},
	urldate = {2025-05-31},
	journal = {Physical Review Letters},
	author = {Haldane, F. D. M.},
	month = oct,
	year = {1988},
	pages = {2015--2018},
}

@article{kane_quantum_2005,
	title = {Quantum {Spin} {Hall} {Effect} in {Graphene}},
	volume = {95},
	url = {https://link.aps.org/doi/10.1103/PhysRevLett.95.226801},
	doi = {10.1103/PhysRevLett.95.226801},
	number = {22},
	urldate = {2025-05-31},
	journal = {Physical Review Letters},
	publisher = {American Physical Society},
	author = {Kane, C. L. and Mele, E. J.},
	month = nov,
	year = {2005},
	pages = {226801},
}

@article{lee_topolectrical_2018,
	title = {Topolectrical {Circuits}},
	volume = {1},
	copyright = {2018 The Author(s)},
	issn = {2399-3650},
	url = {https://www.nature.com/articles/s42005-018-0035-2},
	doi = {10.1038/s42005-018-0035-2},
	language = {english},
	number = {1},
	urldate = {2025-09-04},
	journal = {Communications Physics},
	publisher = {Nature Publishing Group},
	author = {Lee, Ching Hua and Imhof, Stefan and Berger, Christian and Bayer, Florian and Brehm, Johannes and Molenkamp, Laurens W. and Kiessling, Tobias and Thomale, Ronny},
	month = jul,
	year = {2018},
	pages = {39},
}

@article{tapia-de-la-rosa_nanostructured_2024,
	title = {Nanostructured pumped-corner state in a kagome lattice},
	volume = {109},
	issn = {2469-9950, 2469-9969},
	url = {https://link.aps.org/doi/10.1103/PhysRevB.109.085402},
	doi = {10.1103/PhysRevB.109.085402},
	language = {english},
	number = {8},
	urldate = {2026-01-27},
	journal = {Physical Review B},
	author = {Tapia-de-la-Rosa, A. Jazmín and Barrios-Vargas, J. Eduardo},
	month = feb,
	year = {2024},
	pages = {085402},
}

@article{diaz-bonifaz_band_2025,
	title = {Band engineering aided by topological edge state proximity effects: {Inducing} antichirality in graphene},
	volume = {112},
	issn = {2469-9950, 2469-9969},
	shorttitle = {Band engineering aided by topological edge state proximity effects},
	url = {https://link.aps.org/doi/10.1103/9qf8-yfkl},
	doi = {10.1103/9qf8-yfkl},
	language = {english},
	number = {19},
	urldate = {2026-01-27},
	journal = {Physical Review B},
	author = {Díaz-Bonifaz, Ricardo Y. and Ramírez, Carlos},
	month = nov,
	year = {2025},
	pages = {195421},
}

@article{su_solitons_1979,
	title = {Solitons in {Polyacetylene}},
	volume = {42},
	copyright = {http://link.aps.org/licenses/aps-default-license},
	issn = {0031-9007},
	url = {https://link.aps.org/doi/10.1103/PhysRevLett.42.1698},
	doi = {10.1103/PhysRevLett.42.1698},
	language = {english},
	number = {25},
	urldate = {2026-01-27},
	journal = {Physical Review Letters},
	author = {Su, W. P. and Schrieffer, J. R. and Heeger, A. J.},
	month = jun,
	year = {1979},
	pages = {1698--1701},
}

@article{cerjan_observation_2020,
	title = {Observation of a {Higher}-{Order} {Topological} {Bound} {State} in the {Continuum}},
	volume = {125},
	issn = {0031-9007, 1079-7114},
	url = {https://link.aps.org/doi/10.1103/PhysRevLett.125.213901},
	doi = {10.1103/PhysRevLett.125.213901},
	language = {english},
	number = {21},
	urldate = {2026-01-27},
	journal = {Physical Review Letters},
	author = {Cerjan, Alexander and Jürgensen, Marius and Benalcazar, Wladimir A. and Mukherjee, Sebabrata and Rechtsman, Mikael C.},
	month = nov,
	year = {2020},
	pages = {213901},
}

@article{kitaev_unpaired_2001,
	title = {Unpaired {Majorana} fermions in quantum wires},
	volume = {44},
	issn = {1063-7869},
	url = {https://doi.org/10.1070/1063-7869/44/10S/S29},
	doi = {10.1070/1063-7869/44/10S/S29},
	language = {english},
	number = {10S},
	urldate = {2026-01-27},
	journal = {Physics-Uspekhi},
	author = {Kitaev, A. Yu},
	month = oct,
	year = {2001},
	pages = {131},
}

@article{munoz_topological_2018,
	title = {Topological properties of a bipartite lattice of domain wall states},
	volume = {8},
	copyright = {2018 The Author(s)},
	issn = {2045-2322},
	url = {https://www.nature.com/articles/s41598-018-35651-6},
	doi = {10.1038/s41598-018-35651-6},
	language = {english},
	number = {1},
	urldate = {2026-01-28},
	journal = {Scientific Reports},
	publisher = {Nature Publishing Group},
	author = {Munoz, F. and Pinilla, Fernanda and Mella, J. and Molina, Mario I.},
	month = nov,
	year = {2018},
	pages = {17330},
}

@article{longhi_topological_2019,
	title = {Topological pumping of edge states via adiabatic passage},
	volume = {99},
	issn = {2469-9950, 2469-9969},
	url = {https://link.aps.org/doi/10.1103/PhysRevB.99.155150},
	doi = {10.1103/PhysRevB.99.155150},
	language = {english},
	number = {15},
	urldate = {2026-01-28},
	journal = {Physical Review B},
	author = {Longhi, Stefano},
	month = apr,
	year = {2019},
	pages = {155150},
}

@article{english_topologically_2026,
	title = {Topologically protected spatially localized modes: {An} easy experimental realization of the {Su}–{Schrieffer}–{Heeger} model},
	volume = {734},
	issn = {09214526},
	shorttitle = {Topologically protected spatially localized modes},
	url = {https://linkinghub.elsevier.com/retrieve/pii/S0921452626003741},
	doi = {10.1016/j.physb.2026.418616},
	language = {english},
	urldate = {2026-09-21},
	journal = {Physica B: Condensed Matter},
	author = {English, L.Q. and Halchenko, A. and Palmero, F.},
	month = jul,
	year = {2026},
	pages = {418616},
}

@book{asboth_short_2016,
	address = {Cham},
	series = {Lecture {Notes} in {Physics}},
	title = {A {Short} {Course} on {Topological} {Insulators}},
	volume = {919},
	copyright = {http://www.springer.com/tdm},
	isbn = {978-3-319-25605-4 978-3-319-25607-8},
	url = {http://link.springer.com/10.1007/978-3-319-25607-8},
	doi = {10.1007/978-3-319-25607-8},
	urldate = {2026-09-22},
	publisher = {Springer International Publishing},
	author = {Asbóth, János K. and Oroszlány, László and Pályi, András},
	year = {2016},
}

@article{bomantara_quantum_2018,
	title = {Quantum computation via {Floquet} topological edge modes},
	volume = {98},
	url = {https://link.aps.org/doi/10.1103/PhysRevB.98.165421},
	doi = {10.1103/PhysRevB.98.165421},
	number = {16},
	urldate = {2026-10-06},
	journal = {Phys. Rev. B},
	publisher = {American Physical Society},
	author = {Bomantara, Raditya Weda and Gong, Jiangbin},
	month = oct,
	year = {2018},
	pages = {165421},
}

@article{petropoulos_topological_2022,
	title = {Topological order detection and qubit encoding in {Su}–{Schrieffer}–{Heeger} type quantum dot arrays},
	volume = {131},
	issn = {0021-8979},
	url = {https://doi.org/10.1063/5.0082214},
	doi = {10.1063/5.0082214},
	number = {7},
	urldate = {2026-10-06},
	journal = {J. Appl. Phys.},
	author = {Petropoulos, Nikolaos and Bogdan Staszewski, Robert and Leipold, Dirk and Blokhina, Elena},
	month = feb,
	year = {2022},
	pages = {074401},
}

@article{lang_topological_2017,
	title = {Topological networks for quantum communication between distant qubits},
	volume = {3},
	copyright = {2017 The Author(s)},
	issn = {2056-6387},
	url = {https://www.nature.com/articles/s41534-017-0047-x},
	doi = {10.1038/s41534-017-0047-x},
	language = {english},
	number = {1},
	urldate = {2026-10-07},
	journal = {npj Quantum Inf},
	publisher = {Nature Publishing Group},
	author = {Lang, Nicolai and Büchler, Hans Peter},
	month = nov,
	year = {2017},
	pages = {47},
}

@article{kim_quantum_2021,
	title = {Quantum {Electrodynamics} in a {Topological} {Waveguide}},
	volume = {11},
	issn = {2160-3308},
	url = {https://link.aps.org/doi/10.1103/PhysRevX.11.011015},
	doi = {10.1103/PhysRevX.11.011015},
	language = {english},
	number = {1},
	urldate = {2026-10-07},
	journal = {Phys. Rev. X},
	author = {Kim, Eunjong and Zhang, Xueyue and Ferreira, Vinicius S. and Banker, Jash and Iverson, Joseph K. and Sipahigil, Alp and Bello, Miguel and González-Tudela, Alejandro and Mirhosseini, Mohammad and Painter, Oskar},
	month = jan,
	year = {2021},
	pages = {011015},
}

@article{kong_opportunities_2011,
	title = {Opportunities in chemistry and materials science for topological insulators and their nanostructures},
	volume = {3},
	copyright = {2011 Springer Nature Limited},
	issn = {1755-4349},
	url = {https://www.nature.com/articles/nchem.1171},
	doi = {10.1038/nchem.1171},
	language = {english},
	number = {11},
	urldate = {2026-10-07},
	journal = {Nature Chem},
	publisher = {Nature Publishing Group},
	author = {Kong, Desheng and Cui, Yi},
	month = nov,
	year = {2011},
	pages = {845--849},
}

@article{ginley_topological_2016,
	title = {Topological {Insulator} {Film} {Growth} by {Molecular} {Beam} {Epitaxy}: {A} {Review}},
	volume = {6},
	copyright = {http://creativecommons.org/licenses/by/3.0/},
	issn = {2073-4352},
	shorttitle = {Topological {Insulator} {Film} {Growth} by {Molecular} {Beam} {Epitaxy}},
	url = {https://www.mdpi.com/2073-4352/6/11/154},
	doi = {10.3390/cryst6110154},
	language = {english},
	number = {11},
	urldate = {2026-10-07},
	journal = {Crystals},
	publisher = {Multidisciplinary Digital Publishing Institute},
	author = {Ginley, Theresa P. and Wang, Yong and Law, Stephanie},
	month = nov,
	year = {2016},
	pages = {154},
}

@article{he_review_2013,
	title = {Review of {3D} topological insulator thin-film growth by molecular beam epitaxy and potential applications},
	volume = {7},
	issn = {1862-6270},
	url = {https://onlinelibrary.wiley.com/doi/abs/10.1002/pssr.201307003},
	doi = {10.1002/pssr.201307003},
	language = {english},
	number = {1-2},
	urldate = {2026-10-07},
	journal = {physica status solidi (RRL) – Rapid Research Letters},
	author = {He, Liang and Kou, Xufeng and Wang, Kang L.},
	year = {2013},
	pages = {50--63},
}

@article{aidelsburger_artificial_2018,
	title = {Artificial gauge fields and topology with ultracold atoms in optical lattices},
	volume = {51},
	issn = {0953-4075},
	url = {https://iopscience.iop.org/article/10.1088/1361-6455/aac120},
	doi = {10.1088/1361-6455/aac120},
	language = {english},
	number = {19},
	urldate = {2026-10-07},
	journal = {J. Phys. B: At. Mol. Opt. Phys.},
	publisher = {IOP Publishing},
	author = {Aidelsburger, M.},
	month = sep,
	year = {2018},
	pages = {193001},
}

@article{leder_real-space_2016,
	title = {Real-space imaging of a topologically protected edge state with ultracold atoms in an amplitude-chirped optical lattice},
	volume = {7},
	copyright = {2016 The Author(s)},
	issn = {2041-1723},
	url = {https://www.nature.com/articles/ncomms13112},
	doi = {10.1038/ncomms13112},
	language = {english},
	number = {1},
	urldate = {2026-10-07},
	journal = {Nat Commun},
	publisher = {Nature Publishing Group},
	author = {Leder, Martin and Grossert, Christopher and Sitta, Lukas and Genske, Maximilian and Rosch, Achim and Weitz, Martin},
	month = oct,
	year = {2016},
	pages = {13112},
}

@article{RAMIREZ_SMat,
title = {Scattering matrix of arbitrary tight-binding Hamiltonians},
journal = {Annals of Physics},
volume = {378},
pages = {303-316},
year = {2017},
issn = {0003-4916},
doi = {https://doi.org/10.1016/j.aop.2017.01.015},
url = {https://www.sciencedirect.com/science/article/pii/S0003491617300271},
author = {C. Ramírez and L.A. Medina-Amayo}
}

@book{ryndyk_2016,
	address = {Cham},
	series = {Springer {Series} in {Solid}-{State} {Sciences}},
	title = {Theory of {Quantum} {Transport} at {Nanoscale}: {An} {Introduction}},
	isbn = {978-3-319-24086-2 978-3-319-24088-6},
	url = {http://link.springer.com/10.1007/978-3-319-24088-6},
	doi = {10.1007/978-3-319-24088-6},
	urldate = {2026-10-15},
	publisher = {Springer International Publishing},
	author = {Ryndyk, Dmitry},
	year = {2016}
}

@article{
pn-junctions,
author = {J. R. Williams  and L. DiCarlo  and C. M. Marcus },
title = {Quantum Hall Effect in a Gate-Controlled $p-n$ Junction of Graphene},
journal = {Science},
volume = {317},
number = {5838},
pages = {638-641},
year = {2007},
doi = {10.1126/science.1144657},
URL = {https://www.science.org/doi/abs/10.1126/science.1144657},
eprint = {https://www.science.org/doi/pdf/10.1126/science.1144657}
}

@article{liu_fully_2022,
	title = {Fully integrated topological electronics},
	volume = {12},
	copyright = {2022 The Author(s)},
	issn = {2045-2322},
	url = {https://www.nature.com/articles/s41598-022-17010-8},
	doi = {10.1038/s41598-022-17010-8},
	language = {english},
	number = {1},
	urldate = {2026-10-30},
	journal = {Sci Rep},
	publisher = {Nature Publishing Group},
	author = {Liu, Yuqi and Cao, Weidong and Chen, Weijian and Wang, Hua and Yang, Lan and Zhang, Xuan},
	month = aug,
	year = {2022},
	pages = {13410},
}

@book{datta1995electronic,
  title     = {Electronic Transport in Mesoscopic Systems},
  author    = {Datta, Supriyo},
  series    = {Cambridge Studies in Semiconductor Physics and Microelectronic Engineering},
  year      = {1995},
  publisher = {Cambridge University Press},
  address   = {Cambridge},
  doi       = {10.1017/CBO9780511805776},
  isbn      = {9780521416474}
}

@article{Benalcazar,
  title = {Generalized bulk-boundary correspondence in non-Hermitian lattices: Driven circuit realization},
  author = {Helbig, T. and Hofmann, T. and Imhof, S. and Abdelghany, M. and Kiessling, T. and Molenkamp, L. W. and Lee, C. H. and Szameit, A. and Greiter, M. and Thomale, R.},
  journal = {Phys. Rev. B},
  volume = {101},
  issue = {16},
  pages = {161116},
  numpages = {6},
  year = {2020},
  month = {Apr},
  publisher = {American Physical Society},
  doi = {10.1103/PhysRevB.101.161116},
  url = {https://link.aps.org/doi/10.1103/PhysRevB.101.161116}
}

@article{Farchioni1992,
  author  = {Farchioni, Riccardo and Grosso, Giuseppe and Pastori Parravicini, Giuseppe},
  title   = {Electronic structure in incommensurate potentials obtained using a numerically accurate renormalization scheme},
  journal = {Physical Review B},
  volume  = {49},
  pages   = {6383},
  year    = {1992},
  doi     = {10.1103/PhysRevB.45.6383}
}

\end{document}